\documentclass[conference]{IEEEtran}
\IEEEoverridecommandlockouts
\usepackage{cite}
\usepackage{amsmath,amssymb,amsfonts}
\usepackage{algorithmic}
\usepackage{graphicx}
\usepackage{textcomp}
\usepackage{xcolor}
\usepackage{hyperref}

\hypersetup{
    colorlinks=true, 
    linkcolor=black,  
    citecolor=black, 
    filecolor=black, 
    urlcolor=black    
}

\def\BibTeX{{\rm B\kern-.05em{\sc i\kern-.025em b}\kern-.08em
    T\kern-.1667em\lower.7ex\hbox{E}\kern-.125emX}}
\begin{document}

\title{Quantum-Trained Convolutional Neural Network for Deepfake Audio Detection
\thanks{* Corresponding Author: kuan-cheng.chen17@imperial.ac.uk\\
The views expressed in this article are those of the authors and do not represent the views of Wells Fargo. This article is for informational purposes only. Nothing contained in this article should be construed as investment advice. Wells Fargo makes no express or implied warranties and expressly disclaims all legal, tax, and accounting implications related to this article.}
}

\author{
\IEEEauthorblockN{
    Chu-Hsuan Abraham Lin\IEEEauthorrefmark{2}\IEEEauthorrefmark{3},
    Chen-Yu Liu\IEEEauthorrefmark{4}, 
    Samuel Yen-Chi Chen\IEEEauthorrefmark{5}, 
    Kuan-Cheng Chen\IEEEauthorrefmark{2}\IEEEauthorrefmark{3}\IEEEauthorrefmark{1}
}

\IEEEauthorblockA{\IEEEauthorrefmark{2}Centre for Quantum Engineering, Science and Technology (QuEST), Imperial College London, London, UK}
\IEEEauthorblockA{\IEEEauthorrefmark{3}Department of Electrical and Electronic Engineering, Imperial College London, London, UK}
\IEEEauthorblockA{\IEEEauthorrefmark{4}Graduate Institute of Applied Physics, National Taiwan University, Taipei, Taiwan}
\IEEEauthorblockA{\IEEEauthorrefmark{5}Wells Fargo, New York, NY, USA}
}

\maketitle

\begin{abstract}
The rise of deepfake technologies has posed significant challenges to privacy, security, and information integrity, particularly in audio and multimedia content. This paper introduces a Quantum-Trained Convolutional Neural Network (QT-CNN) framework designed to enhance the detection of deepfake audio, leveraging the computational power of quantum machine learning (QML). The QT-CNN employs a hybrid quantum-classical approach, integrating Quantum Neural Networks (QNNs) with classical neural architectures to optimize training efficiency while reducing the number of trainable parameters. Our method incorporates a novel quantum-to-classical parameter mapping that effectively utilizes quantum states to enhance the expressive power of the model, achieving up to 70\% parameter reduction compared to classical models without compromising accuracy. Data pre-processing involved extracting essential audio features, label encoding, feature scaling, and constructing sequential datasets for robust model evaluation. Experimental results demonstrate that the QT-CNN achieves comparable performance to traditional CNNs, maintaining high accuracy during training and testing phases across varying configurations of QNN blocks. The QT framework’s ability to reduce computational overhead while maintaining performance underscores its potential for real-world applications in deepfake detection and other resource-constrained scenarios. This work highlights the practical benefits of integrating quantum computing into artificial intelligence, offering a scalable and efficient approach to advancing deepfake detection technologies.
\end{abstract}


\section{Introduction}
The rapid advancement of machine learning (ML) has paved the way for innovative applications, but it has also introduced challenges, particularly in the realm of synthetic media, such as deepfake audio and video\cite{westerlund2019emergence}. Deepfakes, which leverage deep learning techniques to create highly realistic yet counterfeit multimedia content, pose significant threats to privacy, security, and information integrity\cite{pantserev2020malicious}. As deepfakes become increasingly sophisticated, detecting these manipulations has become a critical area of research. Traditional detection methods \cite{hamza2022deepfake, dixit2023review} rely on classical machine learning models; however, these approaches often struggle with generalization and scalability when confronted with the latest generative adversarial networks (GANs) and advanced deep learning architectures that are used to create deepfakes\cite{guarnera2024mastering}.

To address these challenges, our research explores the use of Quantum Machine Learning (QML) to enhance the detection capabilities of convolutional neural networks (CNNs) specifically for deepfake audio. QML, an emerging field that combines the computational power of quantum mechanics with the versatility of machine learning, has shown promising potential in various domains\cite{biamonte2017quantum, huang2021power, schuld2019quantum, perez2020data, chen2020variationalQRL, chen2022quantumLSTM, liu2022implementation, chen2024quantum2, liu2023learning, chen2024quantum, chen2024cutnqsvmcutensornetacceleratedquantumsupport, ho2024quantum}. Quantum neural networks (QNNs), for instance, exploit quantum superposition and entanglement to process information in ways that classical networks cannot, potentially accelerating training times and improving performance on complex data sets \cite{abbas2021power}. By leveraging quantum-enhanced techniques, QML can offer unique advantages in scenarios where large-scale data processing and intricate pattern recognition are essential\cite{huang2021power}.

\begin{figure}[!t]
\centering
\includegraphics[scale=0.4]{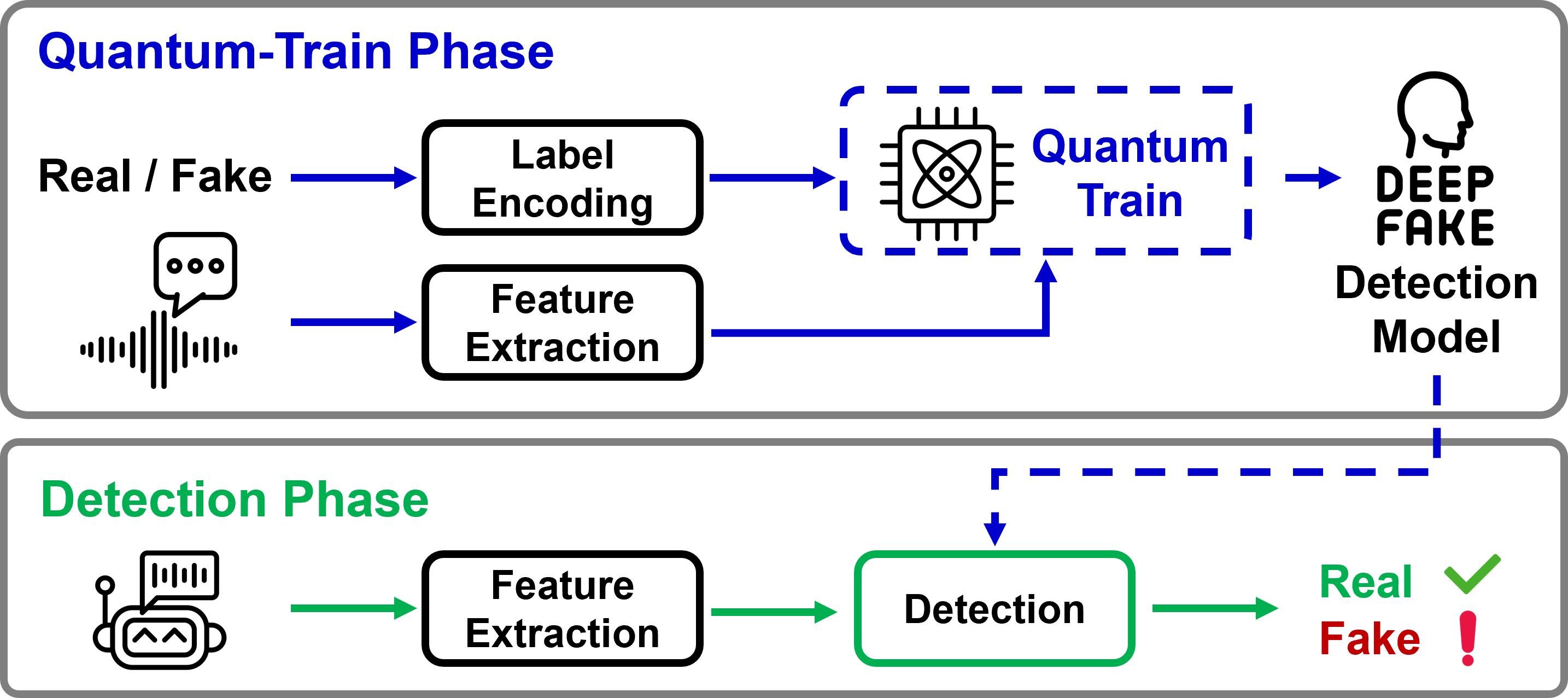}
\caption{Conceptual diagram of the Quantum-Train framework, where the blue line represents the quantum-assisted training of the detection model, and the green line indicates inference on classical hardware such as GPUs for real-time classification of audio as real or fake.
 }
\label{fig:concept}
\end{figure}

This study introduces a Quantum-Trained Convolutional Neural Network (QT-CNN) designed for deepfake audio detection, merging the strengths of quantum computing with classical neural network architectures. Our approach employs a hybrid quantum-classical training mechanism, where quantum circuits are utilized to optimize the weight parameters of the CNN, thus enhancing its performance and robustness against adversarial inputs \cite{west2023towards}. This methodology not only addresses the inherent limitations of classical CNNs in detecting deepfake content but also circumvents some of the scaling issues faced by purely quantum models in the Noisy Intermediate-Scale Quantum (NISQ) era \cite{bharti2022noisy}. By integrating quantum computation in the training phase, our QT-CNN can operate efficiently on classical hardware during the inference stage, making it a practical solution for real-world applications.

In Fig. \ref{fig:concept}, the proposed QT-CNN framework represents a novel intersection of quantum computing and deep learning, tailored to the critical task of detecting deepfake audio. It showcases the practical benefits of quantum-enhanced learning while mitigating the challenges associated with purely quantum models, positioning it as a promising tool in the fight against the proliferation of synthetic media.

\section{Related Works of Quantum Train}
Quantum Machine Learning (QML) has introduced innovative methods to enhance neural network training, particularly through the Quantum-Train (QT) framework \cite{liu2024quantum, liu2024federated, liu2024qtrl, lin2024quantum, liu2024training}, which addresses the escalating complexity and parameter demands of classical neural networks (NNs). QT leverages the expansive Hilbert space of quantum systems to significantly reduce the number of parameters required for training, scaling down from M parameters in classical NNs to \(O(polylog(M))\) by using QNNs\cite{liu2024quantum}. This approach eliminates common challenges faced by traditional QML, such as data encoding complexities and dependency on quantum resources during inference, making the trained models operable on classical hardware. QT has demonstrated its versatility and effectiveness in various applications, including image classification on datasets like MNIST and CIFAR-10 \cite{liu2024quantum,liu2024training}, reinforcement learning \cite{liu2024qtrl}, and time-series forecasting for flood prediction \cite{lin2024quantum}. It further shows potential in federated learning by compressing models and maintaining performance while reducing computational overhead \cite{liu2024federated}. The QT framework, by integrating quantum advantages during training while keeping inference purely classical, offers a practical and scalable solution, making QML more accessible for real-world applications and opening new avenues for the integration of quantum insights into classical machine learning.

\section{Quantum-Train for QNNs}
\label{sec:qt}

\begin{figure*}[!t]
\centering
\includegraphics[scale=0.35]{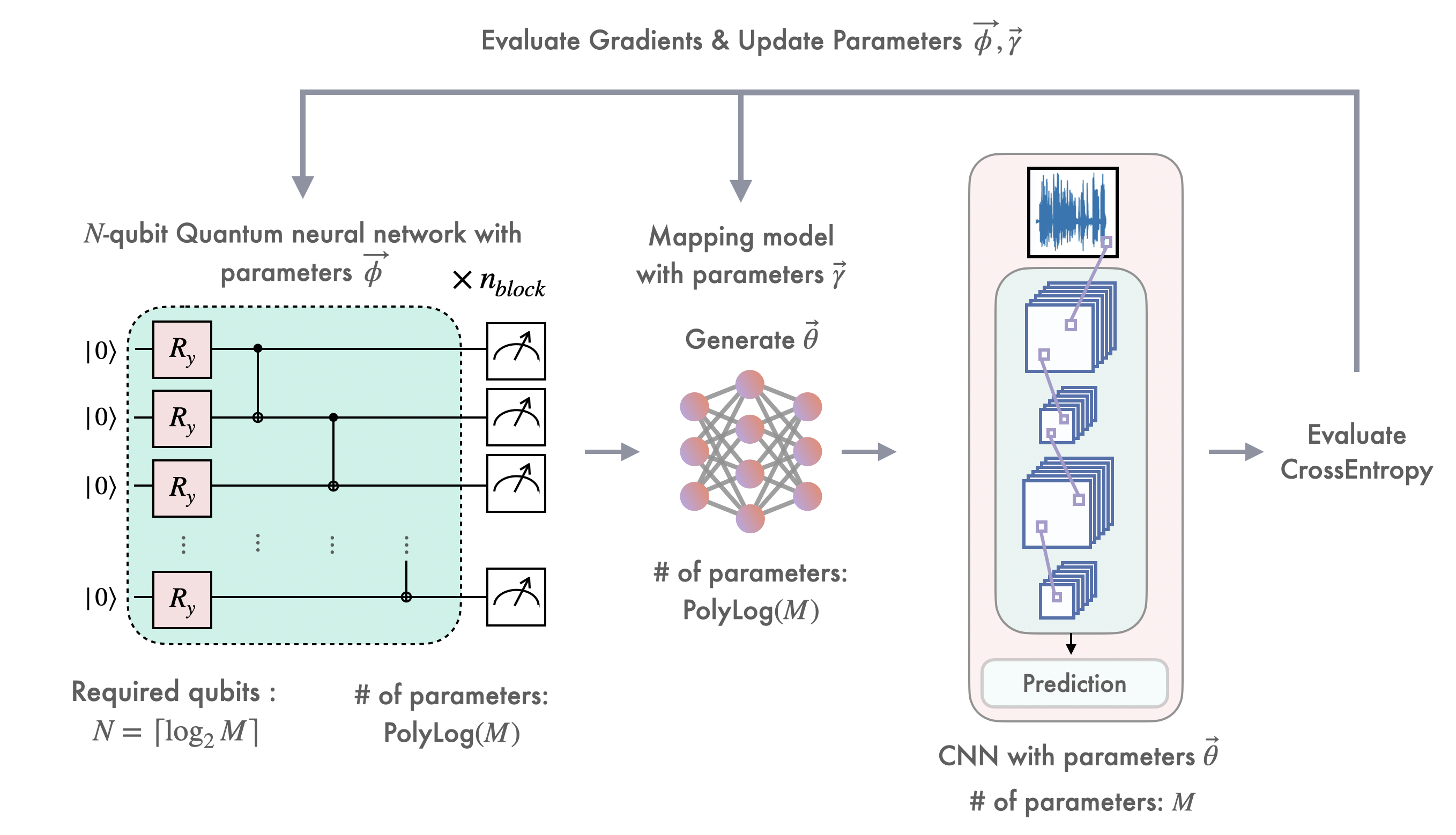}
\caption{Schematic of the QT framework illustrating the flow from the N-qubit QNN with parameterized Ry gates, through the mapping model, to the final CNN. The QT framework significantly reduces the number of trainable parameters by leveraging quantum computation, with gradients evaluated and parameters updated iteratively to optimize the CNN's performance.
}
\label{fig:scheme}
\end{figure*}

\subsection{Quantum-to-Classical Parameter Mapping}

We begin by defining a classical NN characterized by the parameter vector $\vec{\theta} = (\theta_1, \theta_2, \ldots, \theta_M)$. A quantum state $|\psi\rangle$ is encoded using $N = \lceil \log_2 M \rceil$ qubits, creating a Hilbert space of dimension $2^{\lceil \log_2 M \rceil}$ to represent the NN parameters. The measurement probabilities of the quantum state, $|\langle i|\psi\rangle|^2$, range from 0 to 1 for $i \in \{1, 2, \ldots, 2^N\}$. The goal is to map these probabilities to the NN parameters $\theta_i$, which traditionally span from $-\infty$ to $\infty$.

A mapping model $G_{\vec{\gamma}}$ is introduced, built upon an additional NN with parameters $\vec{\gamma}$. It takes input vectors $\vec{x}_i$, such as $[0, 1, 0, 0, 1, 0, 0, 0.023]$ for a 7-qubit system, integrating basis information and measurement probabilities, e.g., $|\langle 0100100|\psi\rangle|^2 = 0.023$. The mapping function $G_{\vec{\gamma}}(\vec{x}_i) = \theta_i$ dynamically determines $\theta_i$, allowing sign flexibility compared to prior models. This adaptation broadens the model’s applicability across diverse machine learning (ML) tasks.

\subsection{Construction of QNNs}
The quantum state $|\psi\rangle$ is constructed with parameterized rotational gates, specifically the Ry gate, represented by:

\begin{equation}
Ry(\mu) = \begin{bmatrix} \cos(\mu/2) & -\sin(\mu/2) \\ \sin(\mu/2) & \cos(\mu/2) \end{bmatrix}.
\end{equation}

Entanglement is introduced via CNOT gates in a linear layout, ensuring parameter scalability as $O(\text{polylog}(M))$. This setup forms the QNN, with parameters $\vec{\phi}$ determining the classical NN weights via the mapping model $G_{\vec{\gamma}}$.

\subsection{Training Procedure for Quantum-Enhanced Learning}
The QT framework involves constructing an $N$-qubit QNN with parameterized Ry gates in blocks, repeatable $n_{\text{block}}$ times to enhance capacity. The classical NN uses parameters $\vec{\theta}$ generated through quantum-classical mapping for traditional training, guided by the cross-entropy loss function:

\begin{equation}
\ell_{\text{CE}} = -\frac{1}{N_d} \sum_{n=1}^{N_d} [y_n \log \hat{y}_n + (1 - y_n) \log (1 - \hat{y}_n)],
\end{equation}

where $y_n$ and $\hat{y}_n$ are the true and predicted labels, respectively. Gradients for the QNN parameters $\vec{\phi}$ and mapping model $\vec{\gamma}$ are computed analytically or via the parameter-shift rule for shot-based simulations. These gradients guide iterative updates, refining the NN parameters.

The QT framework, as depicted in Fig. 1, optimizes efficiency by reducing QNN parameters to $O(\text{polylog}(M))$, compared to $M$ for classical NNs, and supports inference on classical hardware, enhancing practicality under limited quantum resources.

\section{Numerical Results and Discussion}
\label{sec:nrd}

\subsection{Experimental Setup - Deepfake Audio}
The DEEP-VOICE dataset\cite{bird2023real} was employed to evaluate the effectiveness of the QT-CNN in detecting deepfake audio. This dataset is specifically curated to address the rising concerns associated with generative AI technologies that enable voice cloning and real-time voice conversion. DEEP-VOICE \cite{bird2023real} comprises both authentic human speech and AI-generated deepfake audio, designed to simulate realistic scenarios of voice manipulation. The dataset includes a balanced collection of real and synthetic audio samples, providing a comprehensive platform for training and evaluating machine learning models in distinguishing between genuine and manipulated speech. Audio features were systematically extracted and processed to create structured data suitable for model training, testing, and validation. This setup ensures that the QT-CNN model is exposed to diverse and challenging cases of deepfake audio, allowing for robust performance evaluation and benchmarking against conventional detection methods.

\subsection{Data Pre-processing}
The data pre-processing for deepfake audio detection involved preparing the dataset to enhance the performance of the QT-CNN. The dataset consisted of extracted audio features such as chroma STFT \cite{muller2015short}, spectral centroid \cite{gajic2001robust}, spectral bandwidth \cite{iser2008bandwidth}, roll-off frequency, zero-crossing rate \cite{bhattacharjee2018time}, and Mel-Frequency Cepstral Coefficients (MFCCs)\cite{hamza2022deepfake}, which were read into a pandas DataFrame for exploratory data analysis (EDA) \cite{chatfield1986exploratory}. Initial EDA confirmed that the dataset was balanced, containing equal numbers of fake and real audio samples, ensuring unbiased training. Labels were encoded using LabelEncoder to transform categorical labels into numerical values, suitable for machine learning models. Feature scaling was performed using MinMaxScaler to normalize the data, which improves the convergence of neural network models. A sliding window technique was applied to create sequential feature sets, essential for training models. The dataset was then split into training, validation, and test sets using stratified sampling to preserve the balance of classes across all subsets. Correlation matrix analysis was conducted to examine inter-feature relationships, which informed potential feature selection and dimensionality reduction steps. This pre-processing approach ensures that the QT-CNN model is trained on well-prepared, balanced, and normalized data, optimizing its ability to detect deepfake audio accurately.

\begin{figure}[!t]
    \centering
    \includegraphics[width=1\linewidth]{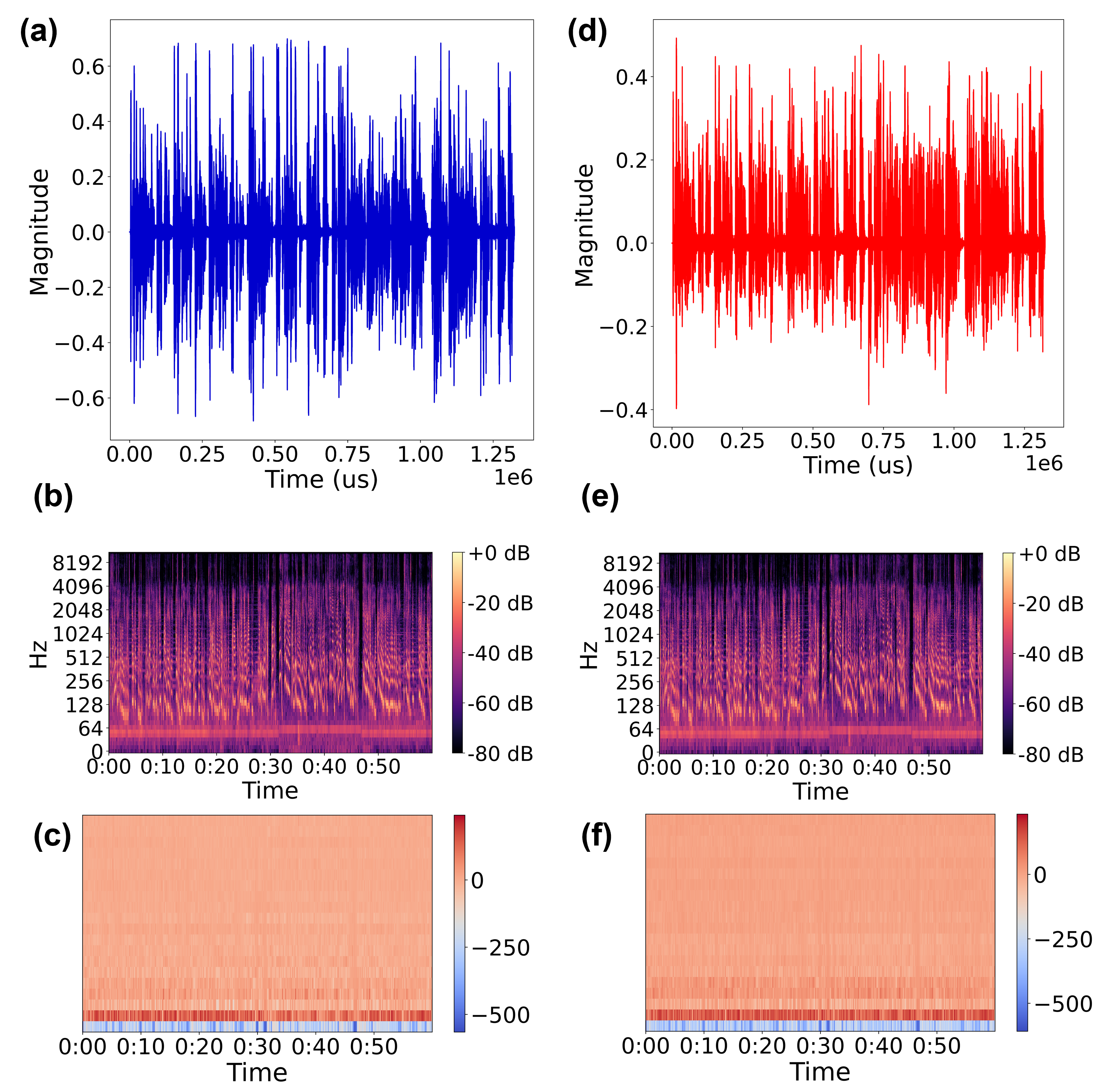}
    \caption{Comparison of real (left: a, b, c) and deepfaked (right: d, e, f) audio representations. Panels (a) and (d) show time-domain waveforms, (b) and (e) display spectrograms, and (c) and (f) present MFCCs, highlighting key audio features.}
    \label{fig:climate-quantum}
\end{figure}

\subsection{Classification Result and Analysis}

\begin{figure}[!b]
\centering
\includegraphics[scale=0.8]{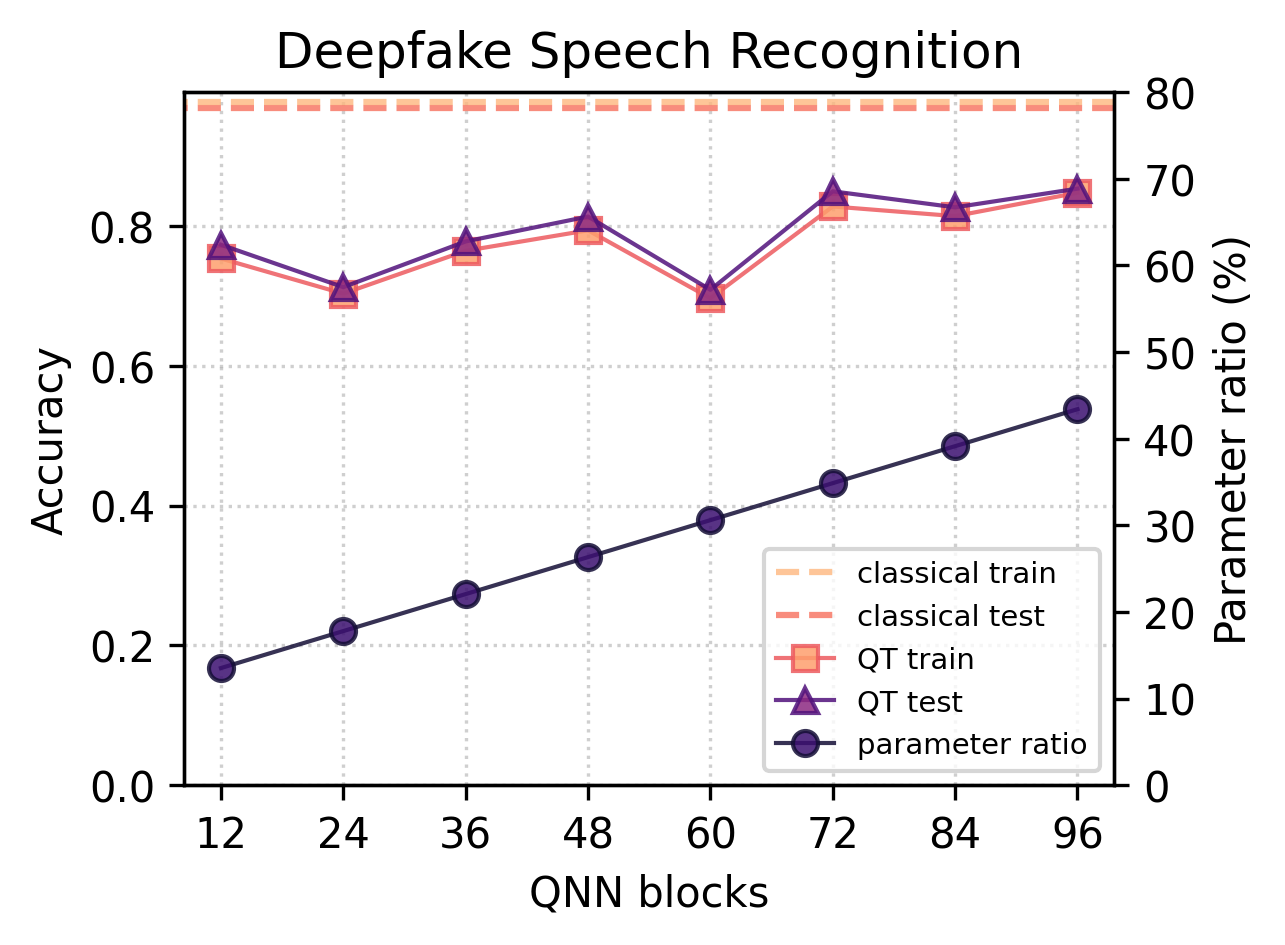}
\caption{Performance comparison of Quantum-Train (QT) CNN and classical CNN models for deepfake speech recognition, demonstrating accuracy and parameter efficiency across varying numbers of QNN blocks. The QT approach achieves comparable accuracy to classical training methods while significantly reducing the parameter count, as indicated by the increasing parameter ratio curve.
 }
\label{fig:result}
\end{figure}

The original CNN model used in our experiments consists of two convolutional layers and two fully connected layers, encompassing a total of 3,373 parameters. To align with the parameter requirements of this classical model, our QNN blocks were designed to include 12 qubits. The Quantum-Train (QT) model incorporates QNN blocks, a mapping model, and a scaling model. The mapping model includes 301 parameters, while the scaling model comprises 8 parameters. The total number of trainable parameters in the QT model varies with the number of QNN blocks, ranging from 456 to 1,464 parameters as the number of QNN blocks increases from 12 to 96. This corresponds to a parameter ratio ranging from 13.5\% to 43.4\% relative to the classical model, demonstrating the QT model’s ability to significantly reduce the parameter count while still maintaining effective performance.

The performance comparison between the QT-CNN method and the classical training approach for Deepfake Speech Recognition highlights the strengths of the QT framework. The results indicate that QT maintains consistently high accuracy during both training and testing phases across various QNN block configurations. Despite the reduction in the number of trainable parameters, QT’s performance closely aligns with that of classical models, showcasing its robustness and reliability. As the number of QNN blocks increases, the QT model preserves accuracy with only minor deviations from classical benchmarks, reinforcing its capability to achieve parameter reduction without a substantial compromise in performance. The secondary y-axis of the analysis figure depicts the parameter ratio, revealing a clear trend of increased parameter efficiency as the number of QNN blocks grows, reaching up to 70\% efficiency at 96 blocks. This substantial reduction in parameters while maintaining competitive performance is particularly advantageous in resource-constrained environments, such as edge computing or real-time processing scenarios. The results validate the QT framework as a scalable, resource-efficient alternative to traditional CNN training, positioning it as a promising tool for advancing hybrid quantum-classical models in AI and quantum computing applications.

\section{Conclusion and Future Work}
\label{sec:cfw}

The results of this study underscore the significant potential of the QT framework in enhancing the performance and efficiency of neural networks, particularly in the context of deepfake audio detection. By integrating quantum computation during the training phase, the QT framework demonstrates the capability to significantly reduce the number of trainable parameters required by classical CNNs, without compromising on performance. The experiments reveal that QT maintains accuracy levels comparable to traditional methods across varying configurations of QNN blocks, even as it reduces the parameter count by up to 70\%. This reduction is achieved through the unique parameter mapping and distributed circuit design, which leverage quantum properties to optimize weight generation for the CNN. These findings highlight the framework’s ability to address critical scalability and resource constraints, making quantum-enhanced learning an attractive approach for tackling complex machine learning tasks, particularly where conventional methods fall short.

In conclusion, this work presents the QT-CNN framework as a novel and practical solution for enhancing deepfake detection and other applications requiring high computational efficiency and robustness. The distributed quantum-classical training approach not only circumvents the limitations of purely quantum models in the NISQ era but also extends the applicability of quantum-enhanced learning to real-world scenarios where inference can be performed on classical hardware. This capability is particularly valuable in domains such as climate change mitigation\cite{lin2024quantum,ho2024quantum}, sustainable development\cite{ho2024quantum}, and beyond, where advanced machine learning models must operate efficiently under resource constraints. In practical quantum systems, this algorithm can also integrate with quantum error mitigation\cite{chen2023short}, distributed quantum computing\cite{burt2024generalised}, and Resource-efficient management in quantum HPC\cite{chen2024noise}. As quantum computing hardware continues to evolve, the QT framework's approach of parameter reduction and distributed circuit design positions it well for future advancements, driving further integration of quantum insights into classical AI methodologies. This work paves the way for more scalable and accessible quantum machine learning solutions, fostering broader adoption and unlocking new possibilities across diverse technological and scientific fields.

\section*{Acknowledgment}
This work was supported by the Engineering and Physical Sciences Research Council (EPSRC) under grant number EP/W032643/1.

\bibliographystyle{ieeetr}
\bibliography{references,ref_YC}

\end{document}